\documentclass[12pt]{article}

\usepackage[dvips]{graphicx}
\usepackage{epsfig}
\usepackage{amsmath,amsfonts,amssymb,amsthm}
\usepackage{verbatim}
\usepackage{psfrag}
\usepackage{bm}
\usepackage{bbm}
\usepackage[square,comma,sort&compress,numbers]{natbib}
\usepackage{color}
\usepackage{slashed}
\usepackage{upgreek}

\usepackage{epsf,epsfig}
\usepackage{graphics}

\newcommand{\gsim}
{\;\raisebox{-.3em}{$\stackrel{\displaystyle >}{\sim}$}\;}

\begin{document}
\thispagestyle{empty}

\begin{flushright}
{
\small
TTK-12-28
}
\end{flushright}

\vspace{0.4cm}
\begin{center}
\Large\bf\boldmath
Leptogenesis from a GeV Seesaw without Mass Degeneracy
\unboldmath
\end{center}

\vspace{0.4cm}

\begin{center}
{Marco Drewes and Bj\"orn~Garbrecht}\\
\vskip0.2cm
{\it Institut f\"ur Theoretische Teilchenphysik und Kosmologie,\\ 
RWTH Aachen University, 52056 Aachen, Germany}\\
\vskip1.4cm
\end{center}

\begin{abstract}
For Leptogenesis based on the type-I seesaw mechanism,
we present a systematic calculation of lepton-number violating
and purely flavoured asymmetries
within nonequilibrium Quantum Field Theory.
We show that sterile neutrinos with non-degenerate masses in the GeV
range can explain the baryon asymmetry of the Universe via flavoured 
Leptogenesis. This is possible due to the interplay of thermal and
flavour effects. Our approach clarifies the relation between
Leptogenesis from the oscillations of sterile neutrinos and
the more commonly studied scenarios from decays and inverse decays.
We explain why lower mass bounds for non-degenerate sterile neutrinos
derived for Leptogenesis from out-of-equilibrium
decays do not apply to flavoured
Leptogenesis with GeV-scale neutrinos.
\end{abstract}

%\pacs{05.30.-d, 11.30.Fs, 12.60.Fr, 14.60.St, 95.30.Cq, 98.80.-k}

%\maketitle

\section{Introduction}
Right handed (RH) neutrinos can provide simple and elegant explanations for two pieces of observational evidence for Physics beyond the Standard Model (SM), namely neutrino oscillations and the baryon asymmetry of the universe (BAU). On one hand, they can generate neutrino masses via the seesaw mechanism~\cite{Minkowski:1977sc}. On the other hand, their $CP$-violating interactions in the early Universe can generate a matter-antimatter asymmetry amongst leptons, which can be transferred to the baryonic sector by sphalerons \cite{Kuzmin:1985mm}. This process is known as Leptogenesis~\cite{Fukugita:1986hr}.

There exists a large number of Leptogenesis scenarios, see {\it e.g.}~\cite{DiBari:2012fz} for a recent review. One way to classify them is to distinguish how a deviation from thermal equilibrium, essential for baryogenesis~\cite{Sakharov:1967dj}, and the freeze out of the baryon asymmetry are realised.
In most models studied to date, the asymmetry is generated in decays of RH neutrinos. The deviation from thermal equilibrium in this case is due to the expansion of the Universe,
and Leptogenesis happens due to out-of-equilibrium
decays at the time when the interaction rates
of the RH neutrinos become
Maxwell suppressed.
These scenarios are known as thermal Leptogenesis.
Another possibility is that the asymmetry is generated during the thermal production of right handed neutrinos by $CP$-violating oscillations amongst them~\cite{Akhmedov:1998qx}. Referring to its discoverers, this is sometimes referred to as Akhmedov-Rubakov-Smirnov (ARS) scenario.

Both scenarios are based on the same Lagrangian terms
\begin{align}
\label{Lagrangian}
{\cal L}=\frac{1}{2}\bar\psi_{Ni}({\rm i} \partial\!\!\!/-M_{ij}) \psi_{Nj}
+\bar\psi_{\ell a}{\rm i}\partial\!\!\!/\psi_{\ell a}
+(\partial^\mu\phi^\dagger)(\partial_\mu \phi)
-Y_{ia}^*\bar\psi_{\ell a} \tilde\phi P_{\rm R}\psi_{Ni}
-Y_{ia}\bar\psi_{Ni}P_{\rm L}\phi\psi_{\ell a}\,.
\end{align}
Here $\phi$ is the Higgs doublet, $\tilde\phi=(\epsilon\phi)^\dagger$  ($\epsilon$ is the antisymmetric rank two ${\rm SU}(2)_{\rm L}$ tensor).
The Majorana spinors $\psi_{Ni}$ represent RH neutrinos, and $\psi_{\ell a}$ are the active leptons. We choose a flavour basis where the Majorana mass $M_{ij}$ and charged lepton Yukawa matrices are diagonal.
Besides the Yukawa couplings $Y_{ia}$, interactions with active right-handed leptons are of importance, because these lead to the decoherence of off-diagonal correlations of $\ell_a$, with $a=e,\mu,\tau$. At the temperatures that the present work is concerned with, the decoherence can be assumed as complete~\cite{Endoh:2003mz,Pilaftsis:2005rv,Abada:2006fw,Nardi:2006fx}. For a discussion of the case of incomplete flavour decoherence, {\it cf.} Ref.~\cite{Beneke:2010dz}.

Though flavour effects may be important~\cite{Endoh:2003mz,Pilaftsis:2005rv,Abada:2006fw,Nardi:2006fx,Blanchet:2006be}, the matter-antimatter asymmetry in thermal Leptogenesis is usually due to the violation of total lepton number by the Majorana masses $M_{ii}$ of the right handed neutrinos. Since lepton-number violation
is strongly suppressed at low temperatures $T\ll M_{ii}$, the lepton asymmetry can be preserved from washout after the RH neutrino freeze-out though all other fields are in thermal equilibrium.
An important phenomenological consequence for models of this type is that the mass of the lightest sterile neutrino should be larger than $10^9\,{\rm GeV}$~\cite{Davidson:2002qv,Buchmuller:2002rq}. 
Due account of the effects from active lepton flavours~\cite{Endoh:2003mz,Pilaftsis:2005rv,Abada:2006fw,Nardi:2006fx} leads to some changes in the details of the mass bound, including the possibility that the sterile neutrino whose decay leaves behind the lepton asymmetry is not necessarily the lightest~\cite{DiBari:2005st,Engelhard:2006yg,Antusch:2010ms}. They, however, have no qualitative impact on the mass bound, which remains far above the Electroweak scale. The large required Majorana masses, which make most model parameters inaccessible to laboratory experiments, are the main disadvantage of thermal Leptogenesis models.

A popular way to evade the lower mass bound on $M_{ii}$ of $10^9\,{\rm GeV}$~\cite{Davidson:2002qv,Buchmuller:2002rq} within models specified by the Lagrangian~(\ref{Lagrangian}) is to introduce a symmetry or a tuning that implies a mass degeneracy \begin{align}
\label{enhancement}
\frac{M_{ii}M_{jj}}{|M_{ii}^2-M_{jj}^2|}\gg1\,.
\end{align}
This option is usually referred to as resonant Leptogenesis~\cite{Pilaftsis:2005rv,Covi:1996wh,Flanz:1996fb,Pilaftsis:1997dr,Pilaftsis:1997jf,Pilaftsis:2003gt}.
The numerator above and correspondingly in the source
for the lepton asymmetry, Eq.~(\ref{S:unflavoured}) below, occurs because the creation of a lepton asymmetry demands an insertion of a lepton-number violating (or, equivalently, helicity reversing) Majorana mass into the loop diagram that gives the main contribution to
the decay asymmetry of the sterile neutrinos.
 
In ARS models of Leptogenesis, on the other hand, it is essential that
not all of the sterile RH neutrinos reach thermal equilibrium  before the sphaleron freeze-out at $T= T_{\rm EW}\sim140$ GeV. Therefore, their Yukawa couplings must not be too large. This constrains the masses $M_{ii}$ from above, as these are related to the Yukawa couplings by the seesaw relation $m_\nu\sim Y^2v^2/(\sqrt 2 M)$. Here $v=246\,{\rm GeV}$ is the vacuum expectation value
of the Higgs field and $m_\nu$ the active neutrino mass scale.
As a result, the total lepton number violation in these models is tiny at $T\gtrsim T_{\rm EW}$ for $M_{ii}< T_{\rm EW}$. 
A non-zero baryon number can be generated even for vanishing total lepton number due to flavour effects and because sphalerons only couple to active neutrinos, which can carry a net asymmetry that is compensated for in the sterile sector. 

The smaller masses make it possible to search for the right handed neutrinos in collider experiments~\cite{Gorbunov:2007ak}. However, the small production rates for sterile neutrinos, essential to avoid their equilibration and washout of the asymmetries long before sphaleron freeze-out, also enter the source term for the asymmetries. Most studies of ARS-type models to date~\cite{Akhmedov:1998qx,Asaka:2005pn,Shaposhnikov:2008pf,Canetti:2010aw,Canetti:2012vf,Canetti:2012zc} require a mass degeneracy of the type~(\ref{enhancement}) to compensate for the resulting suppression. If not justified by a symmetry, this can be considered as tuning.

In this work we show that, due to thermal effects, no mass degeneracy needs to be required to produce the observed BAU.
While the total lepton number violation is always proportional to $M_{ii}$, and hence suppressed unless a degeneracy~(\ref{enhancement}) is
imposed, there is no such suppression for the generation of asymmetries in individual flavours. 
This is a pure medium effect; in vacuum, Lorentz invariance and dimensional arguments imply that the purely flavoured asymmetry is proportional to the mass-square of the decaying sterile neutrino, which has been verified by explicit calculations~\cite{Covi:1996wh,Endoh:2003mz,Nardi:2006fx}. At temperatures $T\gg M_{ii}$ different medium effects are very important\footnote{The rest frame of the heat bath explicitly breaks Lorentz invariance, which generically leads to an enhancement of the numerator term. Moreover, kinematic cuts that contribute to the charge-parity ($CP$) asymmetry, which are suppressed or forbidden in the vacuum, can become important at finite temperature\cite{Garbrecht:2010sz}.}, and one can readily estimate the enhancement factor for the flavoured asymmetries to be
\begin{equation}
\label{enhancement:flavoured}
\frac{T^2}{|M_{ii}^2-M_{jj}^2|}.
\end{equation}
We show explicitly that due to this effect, the flavoured
asymmetries in ARS scenarios are largely enhanced compared to
the total lepton asymmetry.
This may be interpreted a way to circumvent the mass bounds~\cite{Davidson:2002qv,Buchmuller:2002rq}, that apply to the
non-resonant creation of a lepton-number violating asymmetry.
We derive simple analytic expressions
for the flavoured asymmetries,
that allow us to find parameters for which
the observed BAU arises from sterile neutrinos with GeV masses
and without the mass degeneracy~(\ref{enhancement}).
The flavoured asymmetries are effectively converted into a baryon 
asymmetry due to sphaleron conversion, because for our parametric 
examples, the active lepton flavour $e$ is only weakly washed
out before sphaleron freeze-out, while $\mu$ and $\tau$ suffer
a strong washout.
We expect that similar effects may also reduce the lower mass bound or required degree of mass degeneracy for flavoured scenarios in which the asymmetry is generated from heavy neutrino decays.

The relevant temperature in the term~(\ref{enhancement:flavoured}) is
determined by the time that is required to build up the off-diagonal correlations between the sterile neutrinos~\cite{Akhmedov:1998qx}
(or, in other words, the oscillation time of the sterile neutrinos). For ${\rm GeV}$-scale neutrinos with mass-splittings of a similar size, it turns out that $T={\cal O}(10^6\,{\rm GeV})$. Further, we can estimate that the shorter Hubble time $H^{-1}\sim m_{\rm Pl}/T^2$ ($H$ is the Hubble rate and $m_{\rm Pl}=1.22\times 10^{19}\,{\rm GeV}$ is the Planck mass) suppresses the production of the flavoured asymmetries at high temperatures compared to the lepton-number violating asymmetry, that is mostly produced close to sphaleron freeze-out at $T\sim T_{\rm EW}$, by a factor of $T_{\rm EW}^2/T^2$. 
On the other hand, the production rate of sterile neutrinos with $M_{ii}\ll T$ is proportional to $T$, what leads to a relative enhancement $T/T_{\rm EW}$. Combining these enhancements and suppressions with the factor~(\ref{enhancement:flavoured}), one can therefore estimate that the flavoured asymmetries are enhanced by the large factor $T T_{\rm EW}/M_{ii}^2$ when compared to the unflavoured total lepton asymmetry.
A similar enhancement mechanism, that relies on the ratio of the mass of a heavy sterile neutrino much above the Electroweak scale and the mass splitting between several Electroweak-scale Higgs doublets was recently proposed in Ref.~\cite{Garbrecht:2012qv}.

Our results are inferred from the source term~(\ref{Source:flavoured:2}) that is a straightforward generalisation of the results of Ref.~\cite{Garbrecht:2011aw} ({\it cf.} also Ref.~\cite{Garny:2011hg}). The source term is calculated within the framework of nonequilibirum Quantum Field Theory, as described by the Schwinger-Keldysh or Closed-Time-Path (CTP) formalism \cite{Schwinger:1960qe}. Compared to calculations involving vacuum $S$-matrix elements~\cite{Pilaftsis:2005rv,Covi:1996wh,Flanz:1996fb,Pilaftsis:1997dr,Pilaftsis:1997jf,Pilaftsis:2003gt}, CTP methods~\cite{Beneke:2010dz,Garbrecht:2012qv,Garbrecht:2011aw,Garny:2011hg,Buchmuller:2000nd,De Simone:2007rw,Garny:2009rv,Garny:2009qn,Anisimov:2010aq,Garny:2010nj,Beneke:2010wd,Garny:2010nz,Garbrecht:2010sz,Anisimov:2010dk} allow to systematically include all medium effects in a well-controlled approximation scheme~\cite{Calzetta:1986cq,Prokopec:2003pj,Prokopec:2004ic,Cirigliano:2009yt,Herranen:2010mh,Drewes:2010pf,Cirigliano:2011di,Herranen:2011zg,Fidler:2011yq,Garbrecht:2011xw,Tulin:2012re,Drewes:2012qw}. It turns out~\cite{Garbrecht:2011aw} that the source term~(\ref{Source:flavoured:2}) corresponds to a sufficiently good approximation provided the sterile neutrinos have performed more than half of a flavour oscillation and when the mass splitting is larger than the width of the sterile neutrinos. Taking account of these limitations, the use of the source~(\ref{Source:flavoured:2}) allows for analytical estimates of the asymmetries that perhaps  exhibit the relevant parametric dependences in a more direct manner than numerical studies based on the canonical time-evolution of a density matrix. Moreover, it clarifies the relation
between the ARS mechanism and scenarios of thermal Leptogenesis. 

The plan of this paper is as follows: In Section~\ref{sec:source}, we generalise the result of Ref.~\cite{Garbrecht:2011aw} for the $CP$-violating source term to the case of flavoured Leptogenesis. Assuming ${\rm GeV}$-scale sterile neutrinos, we calculate the flavoured asymmetries present at the time of sphaleron freeze-out in Section~\ref{sec:ass}. The expression for the asymmetry is employed in Section~\ref{sec:par:examples} in order to identify points in parameter space that lead to successful Baryogenesis via Leptogenesis without imposing a mass degeneracy on the sterile neutrinos. In Section~\ref{sec:disc}, we summarise and indicate possible directions of future work.

\section{Flavoured and Unflavoured Source Terms when $|M_{ii}^2-M_{jj}^2|\gg|\mathbf k|\Gamma_{Ni}$}
\label{sec:source}

When $|M_{ii}^2-M_{jj}^2|\gg|\mathbf k|\Gamma_{Ni}$, the oscillations of the sterile neutrinos may be
neglected.\footnote{In Ref.~\cite{Garbrecht:2011aw}, this condition is
written in the form $|M_{ii}-M_{jj}|=\Delta M\gg \Gamma_{Ni}$, which is equivalent when
$|\mathbf k|\sim M_{ii},M_{jj}$ or smaller. For relativistic
sterile neutrinos, that we consider here, 
it generalises to $|M_{ii}^2-M_{jj}^2|\gg|\mathbf k|\Gamma_{Ni}$. With this
replacement, the results of Ref.~\cite{Garbrecht:2011aw} also apply to the relativistic case. Note that an enhancement of the asymmetry
for $|M_{ii}^2-M_{jj}^2|\gg|\mathbf k|\Gamma_{Ni}$ is also
reported in Ref.~\cite{Canetti:2012vf}, where numerical solutions to the
canonical
time-evolution of a density matrix are employed.}
Here, $\Gamma_{Ni}$ is the relaxation rate of the non-equilibrium
sterile neutrino, that is the source for the asymmetry, toward equilibrium, and
$|\mathbf k|$ is a typical momentum scale, {\it i.e.} $|\mathbf k|\sim T$, where
$T$ is the background temperature.
In that limit,
Refs.~\cite{Beneke:2010wd,Garbrecht:2011aw} provide compact expressions for the
resonantly enhanced source term of the total lepton asymmetry
for a general spectral (absorptive) self-energy $\slashed\Sigma_N^{\cal A}$,
that can in particular include finite-density effects:
\begin{align}
\label{S:unflavoured}
S=
\sum_{\overset{i,j}{i\not=j}}
\int\frac{d^3 k}{(2\pi)^3}\frac{1}{2\sqrt{|\mathbf k|^2+M_{ii}^2}}
8{\rm i}[{Y_i^*}^2 Y_j^2-Y_i^2 {Y_j^*}^2]
\frac{M_{ii} M_{jj}}{M_{ii}^2-M_{jj}^2} g_w
\hat\Sigma^{\cal A}_{N\mu}\hat\Sigma_N^{{\cal A}\mu}\delta f_{Nii}(\mathbf k)
\,,
\end{align}
where the coupling constants are factored from the spectral-self energies,
\begin{align}
\label{SigmaLR}
\slashed\Sigma_{Nij}^{\cal A}=g_w \gamma_\mu
\sum\limits_a(Y_{ia}^*Y^t_{aj} P_{\rm R} +Y_{ia}Y^\dagger_{aj}P_{\rm L})
\hat{\Sigma}_N^{{\cal A}\mu}\,,
\end{align}
and where $a$ denotes the active lepton flavours in a suitable basis,
that is given by $e,\mu,\tau$ for the present purposes. The
${\rm SU}(2)_{\rm L}$-multiplicity is accounted for by the factor $g_w=2$.
In principle, the left- and right handed contributions differ in the presence
of a lepton asymmetry~\cite{Garbrecht:2011aw}.
Such a backreaction effect may however be neglected within the source term
when the lepton asymmetry is small. The function $\delta f_{Nii}(\mathbf k)$
describes the deviation of $N_i$ from equilibrium.

In order to account for flavoured asymmetries,
we need to modify the expression~(\ref{S:unflavoured}).
The source term then becomes a matrix in the flavour space of active leptons.
The unflavoured expression in Ref.~\cite{Garbrecht:2011aw} is straightforwardly 
generalised to
\begin{align}
\label{Source:flavoured:1}
S_{ab}=-\sum\limits_{\overset{i,j}{i\not=j}}Y_{ia}^* Y_{jb} \int\frac{d^4 k}{(2\pi)^4}
{\rm tr}
\left[
P_{\rm R}{\rm i}\delta S_{N ij}(k) 2P_{\rm L}
\hat{\slashed\Sigma}^{\cal A}_N(k)
\right]\,.
\end{align}
In the fully flavoured regime, which is the concern of the present work, flavour
off diagonal correlations decohere rapidly, and we can readily
delete the off-diagonal components $a\not=b$ for $a,b=e,\mu,\tau$.
The sterile neutrino Wightman function is decomposed as
\begin{align}
\label{helicity:decomposition}
{\rm i}\delta S_N=\sum\limits_{h=\pm}{\rm i}\delta S_{Nh}\,,
\qquad
-{\rm i}\gamma^0 \delta S_{Nh}=\frac14(\mathbbm 1+h \hat k^i \sigma^i)
\otimes \rho^a g_{ah}\,,
\end{align}
and we note that due to the left and right projectors in Eq.~(\ref{Source:flavoured:1}),
only the vectorial and pseudovectorial densities $g_{0h}$ and $g_{3h}$ are
directly relevant for the source.
For $|M_{ii}^2-M_{jj}^2|\gg|\mathbf k|\Gamma_{Ni}$, these can be expressed as
\begin{subequations}
\label{f:nondeg:offdiag}
\begin{align}
\hat g_{0hij}^{{\rm L},{\rm R}}&=
\frac{M_{ii}(M_{ii}+M_{jj})\mp2h|\mathbf k|k^0+2\mathbf k^2}
{k^0 (M_{ii}^2-M_{jj}^2)}
{\rm i}\frac{g_w}{2}\left(
\hat\Sigma^{{\cal A}0}_N
\pm h \hat k^i\hat\Sigma^{{\cal A}i}_N
\right)
g_{0hii}+i\leftrightarrow j\,,
\\
\hat g^{{\rm L},{\rm R}}_{3hij}&=
\frac{M_{ii}(M_{ii}-M_{jj})\mp2h|\mathbf k|k^0+2\mathbf k^2}
{k^0 (M_{ii}^2-M_{jj}^2)}
{\rm i}\frac{g_w}{2}\left(
\mp\hat\Sigma^{{\cal A}0}_N
-h \hat k^i\hat\Sigma^{{\cal A}i}_N
\right)
g_{0hii}+i\leftrightarrow j\,,
\end{align}
\end{subequations}
where [{\it cf.} Eq.~(\ref{SigmaLR})]
\begin{align}
\label{g:offdiag}
g^{\rm L}_{0,3hij}
=
\sum\limits_a
Y_{ia}^* Y^t_{aj} \hat g^{\rm L}_{0,3hij}
\,,
\quad
g^{\rm R}_{0,3hij}
=
\sum\limits_a
Y_{ia} Y^\dagger_{aj} \hat g^{\rm R}_{0,3hij}\,,
\end{align}
and where we may approximate
\begin{align}
g_{0hii}(k)=2\pi\delta(k^2-M_{ii}^2) 2k^0 \delta f_{0hii}\,.
\end{align}
The $ji$ components follow from
$g_{0,3hji}^{\rm L,R}=(g_{0,3hij}^{\rm L,R})^*$.

Substituting these elements into Eq.~(\ref{Source:flavoured:1}) yields
the flavoured source
\begin{align}
\label{Source:flavoured:2}
S_{ab}=&\sum\limits_{\overset{c,i,j}{i\not=j}}
\frac{8{\rm i} g_w}{M_{ii}^2-M_{jj}^2}
\int\frac{d^3 k}{(2\pi)^3 2\sqrt{\mathbf k^2+M_{ii}^2}}
\\\notag
\times&
\Bigg\{
(Y^\dagger_{ai}Y_{ic}Y^\dagger_{cj}Y_{jb}
-Y^t_{ai}Y^*_{ic}Y^t_{cj}Y^*_{jb})
\\\notag
&\hskip2cm\times
\left[
\left(M_{ii}^2+2\mathbf k^2\right)
\left(
{\hat\Sigma_N^{{\cal A}0}}{}^2+{\hat\Sigma_N^{{\cal A}i}}{}^2
\right)
-4|\mathbf k| \sqrt{\mathbf k^2+M_{ii}^2}\hat\Sigma_N^{{\cal A}0}\hat k^i\hat\Sigma_N^{{\cal A}i}
\right]
\\\notag
&
+(Y^\dagger_{ai}Y^*_{ic}Y^t_{cj}Y_{jb}
-Y^t_{ai}Y_{ic}Y^\dagger_{cj}Y^*_{jb})
M_{ii}M_{jj}
\hat\Sigma_{N\mu}^{\cal A}\hat\Sigma_N^{{\cal A}\mu}
\Bigg\}
\times
\delta f_{0 h ii}(\mathbf k)\,.
\end{align}
Note that in the present approximation, we do not need to take account of
the possible
 helicity dependence of the diagonal distribution functions for $N_i$, as
this corresponds to a small backreaction effect, {\it i.e.} we
take $\delta f_{0 + ii}(\mathbf k)=\delta f_{0 - ii}(\mathbf k)$.
(In principle, the partial equilibration of the active
asymmetries with the sterile sector discussed in
Section~\ref{sec:par:examples} may be consistently included through such an account
of backreaction.)
The first term in the curly bracket is the lepton-number conserving, purely flavoured
contribution, while the second term is the lepton-number violating contribution
to the asymmetry. For the ARS mechanism, we discuss in the following
that the purely flavoured term gives the by far dominant contribution
to the baryon asymmetry.
Note that using the cyclicity of the trace, it is easy to
see that the flavour trace of the first term vanishes, as it should.

As a consistency check, notice that in the zero-temperature limit
$\hat\Sigma_N^{{\cal A}\mu}=k^\mu/(32\pi)$, such that
\begin{align}
\label{Sigma:T0}
\left(M_{ii}^2+2\mathbf k^2\right)
\left(
{\hat\Sigma_N^{{\cal A}0}}{}^2+{\hat\Sigma_N^{{\cal A}i}}{}^2
\right)
-4|\mathbf k| k^0 \hat\Sigma_N^{{\cal A}0}\hat k^i\hat\Sigma_N^{{\cal A}i}
=M_{ii}^4/(2^{10}\pi^2)\,,
\end{align}
where $k^0=\sqrt{\mathbf k^2+M_{ii}^2}$ [contributions from
$k^0=-\sqrt{\mathbf k^2+M_{ii}^2}$ are already accounted for in the source
term~(\ref{Source:flavoured:2})].
In that situation, the term~(\ref{Source:flavoured:2}) therefore is manifestly
Lorentz-invariant and reduces to known
results~\cite{Covi:1996wh,Endoh:2003mz,Nardi:2006fx}.

Now, it is important to notice that in weak-washout scenarios (in the
sense that Leptogenesis takes place at temperatures way above $M_{ii}$),
the purely flavoured asymmetry can generically be much enhanced compared to
the lepton-number violating asymmetry, simply because of the fact that for most
of the singlet neutrinos, the momentum is $|\mathbf k|\gg M_{ii}$.
As we are interested in sterile neutrinos of mass below the Electroweak scale,
we may approximate $M_{ii}$ within the numerator terms as {\it zero} and
obtain for the purely flavoured contribution in that limit
\begin{align}
\label{Sigma:Tlarge}
\left(M_{ii}^2+2\mathbf k^2\right)
\left(
{\hat\Sigma_N^{{\cal A}0}}{}^2+{\hat\Sigma_N^{{\cal A}i}}{}^2
\right)
-4|\mathbf k| k^0 \hat\Sigma_N^{{\cal A}0}\hat k^i\hat\Sigma_N^{{\cal A}i}
\approx
2\mathbf k^2\left(\hat\Sigma_N^{{\cal A}0}-
\hat k^i \hat\Sigma_N^{{\cal A}i}\right)^2
\,.
\end{align}
We see that
clearly, the zero-temperature approximation~(\ref{Sigma:T0}) is
not applicable when $|\mathbf k|\sim T \gg M_{ii}$, and it would lead to
a total underestimate of the flavoured asymmetries. This
is the basic reason why the ARS scenario can effectively circumvent
the mass bounds~\cite{Davidson:2002qv,Buchmuller:2002rq}
that apply to the lepton-number violating asymmetries
or to situations where the zero-temperature approximation is valid.

\section{Lepton Asymmetry and Mass Degeneracy}
\label{sec:ass}

Eq.~(\ref{Source:flavoured:2}) exhibits the relevant parametric dependences
on $M_{ii}$, $Y_{ia}$ and $\mathbf k$ in a simple manner. Once
$\hat \Sigma_N^{{\cal A}\mu}(k)$ is known, only a one-dimensional integral remains
to be performed in order to calculate the source term. The calculation
of $\hat \Sigma_N^{{\cal A}\mu}(k)$ for neutrinos, that are light compared to
the temperature, however turns out to be involved. Recent progress on that
problem is reported in Refs.~\cite{Anisimov:2010gy,Besak:2012qm}.
What is calculated is however not $\hat \Sigma_N^{{\cal A}\mu}(k)$ in its
components, but rather
$2 g_w k_\mu \hat\Sigma_N^{{\cal A}\mu}(k)/(2 k^0)$,
the relaxation rate of $\delta f_N(\mathbf k)$
toward equilibrium.

From rescaling by the equilibrium number density of one species
of approximately
massless RH neutrinos, $2 n_N=3/(2\pi^2)\zeta(3)T^3$
(the factor of two accounts for the two helicity states), one may
infer from the rates that are tabulated in
Ref.~\cite{Besak:2012qm} the temperature-averaged
relaxation rate $\Gamma_{\rm av}\approx 3\times 10^{-3} T$
at temperatures of $5\times10^5\,{\rm GeV}$ (it turns out below that
for our choice of parameters, this is the temperature at which the asymmetry
is generated) and $\Gamma_{\rm av}\approx 5\times 10^{-3} T$ at Electroweak
temperatures (which are relevant for the washout),
whereas Ref.~\cite{Asaka:2005pn} employs
$\Gamma_{\rm av}=0.02\times T/8= 2.5\times 10^{-3} T$.
The relaxation rate of the sterile neutrinos is then given by
\begin{align}
\label{gamma:N}
\Gamma_{Ni}=\sum\limits_a Y_{ia}Y^\dagger_{ai}\Gamma_{\rm av}\,.
\end{align}
In what follows,
we parametrise $\Gamma_{\rm av}=\gamma_{\rm av} T$ and
use the more recently and more systematically derived result
of Ref.~\cite{Besak:2012qm}. The averaged relaxation rate corresponds to
\begin{align}
\Gamma_{\rm av}\equiv
\frac{\int\frac{d^3 k}{(2\pi)^3} g_w 2 k_\mu \hat \Sigma_N^{{\cal A}\mu}(k) f_N^{\rm eq}(\mathbf k)/k^0}{\int\frac{d^3 k}
{(2\pi)^3}f_N^{\rm eq}(\mathbf k)}\,.
\end{align}
In the massless limit, where $k^0=|\mathbf k|$, we make the approximation
\begin{subequations}
\label{est:Sigma}
\begin{align}
\label{kslashSigmaslashoverk}
\hat\Sigma_N^{{\cal A}0}-\hat k^i \hat\Sigma_N^{{\cal A}i}&\approx\frac{\Gamma_{\rm av}}{2 g_w}\,,
\\
\hat\Sigma_N^{{\cal A}0}{}^2-\hat\Sigma_N^{{\cal A}i}{}^2&\approx\frac{\Gamma^2_{\rm av}}{4 g_w^2}\,,
\end{align}
\end{subequations}
where in going from the first to the second estimate,
we have used that
$\hat k^i \hat\Sigma_N^{{\cal A}i}\leq \hat\Sigma_N^{{\cal A}0}$,
and we thereby
allow for an ${\cal O}(1)$ underestimation of that contribution, which enters
the lepton-number violating source term. In the following, we
are however mainly interested in the purely flavoured asymmetries, that are
obtained from Eq.~(\ref{kslashSigmaslashoverk}).

Taking for $\delta f_{0hii}(\mathbf k)=-f_N^{\rm eq}(\mathbf k)$ and
decomposing the source into the purely flavoured and the lepton-number
violating contribution, $S=S^{\rm PF}+S^{\rm LNV}$, we may evaluate within
the present approximations
\begin{align}
\label{Source:approximate}
S_{aa}^{\rm LNV}
&=-\sum\limits_{\overset{c}{j\not=i}}
8{\rm i}g_w\frac{Y^\dagger_{ai}Y^*_{ic}Y^t_{cj}Y_{ja}
-Y^t_{ai}Y_{ic}Y^\dagger_{cj}Y^*_{ja}}{M_{ii}^2-M_{jj}^2}
M_{ii}M_{jj}\frac{\Gamma^2_{\rm av}}{4 g_w^2}
\int\limits_0^\infty\frac{|\mathbf k|d|\mathbf k|}{4\pi^2}\frac{1}{{\rm e}^{\beta|\mathbf k|}+1}
\\\notag
&=
-\frac{\rm i}{24 g_w} T^2
\sum\limits_{\overset{c}{j\not=i}}
\frac{Y^\dagger_{ai}Y^*_{ic}Y^t_{cj}Y_{ja}
-Y^t_{ai}Y_{ic}Y^\dagger_{cj}Y^*_{ja}}{M_{ii}^2-M_{jj}^2}
M_{ii}M_{jj}\Gamma^2_{\rm av}
\end{align}
and
\begin{align}
\label{S:PF:int}
S_{aa}^{\rm PF}
&=-\sum\limits_{\overset{c}{j\not=i}}
16{\rm i}g_w\frac{Y^\dagger_{ai}Y_{ic}Y^\dagger_{cj}Y_{ja}
-Y^t_{ai}Y^*_{ic}Y^t_{cj}Y^*_{ja}}{M_{ii}^2-M_{jj}^2}
\frac{\Gamma^2_{\rm av}}{4 g_w^2}
\int\limits_0^\infty\frac{|\mathbf k|^3 d|\mathbf k|}{4\pi^2}\frac{1}{{\rm e}^{\beta|\mathbf k|}+1}
\\\notag
&\approx
-\frac{0.58\,{\rm i}}{g_w}\times T^4
\sum\limits_{\overset{c}{j\not=i}}
\frac{Y^\dagger_{ai}Y_{ic}Y^\dagger_{cj}Y_{ja}
-Y^t_{ai}Y^*_{ic}Y^t_{cj}Y^*_{ja}}{M_{ii}^2-M_{jj}^2}
\Gamma^2_{\rm av}\,.
\end{align}
In principle, there is also a vertex-diagram contributing to the lepton-number
violating source. In the present context of Leptogenesis with ${\rm GeV}$-scale
sterile neutrinos, we may neglect it, since it is not resonantly enhanced and
at the same time suppressed through a helicity-flipping operator.

In the approximation of averaged neutrino production rates, the
kinetic equations are
\begin{subequations}
\label{kineq:eta}
\begin{align}
\frac{d}{d\eta}n_{Ni}
&=-\sum_a Y_{ia} Y_{ai}^\dagger \Gamma_{\rm av} (n_{Ni}-n_N^{\rm eq})\,,
\\
\frac{d}{d\eta}q_{\ell aa}&= g_w S_{aa}\,.
\end{align}
\end{subequations}
Here, $\eta$ denotes the conformal time, $n_{Ni}$ is the number
density of the sterile neutrino $N_i$ of a given helicity
and $q_{aa}$ is the charge density within the lepton flavour $a$,
accounting also for the ${\rm SU}(2)_{\rm L}$-multiplicity.
Then, the temperature $T$ in
$\Gamma_{\rm av}$ and $S$ is understood as a comoving temperature,
that is related to the physical temperature $T_{\rm ph}$ as
$T=a(\eta) T_{\rm ph}$.
In the expressions for the source
terms~(\ref{Source:approximate}), the masses must be conformally
rescaled, $M_{ii}\to a(\eta) M_{ii}$.
In the radiation-dominated Universe,
the scale factor is $a(\eta)=a_{\rm R}\eta$. Here, we choose
\begin{align}
a_{\rm R}=\frac{m_{\rm Pl}}{2}\sqrt{\frac{45}{\pi^3 g_\star}}\,,
\end{align}
where $g_\star=106.75$ is the number of relativistic degrees of freedom
in the symmetric phase of the Standard Model. (The three additional
relativistic sterile neutrinos are assumed to be out-of-equilibrium yet.)
With this choice for $a_{\rm R}$, $T=a_{\rm R}$ and it is useful
to introduce $z=T_{\rm EW}\eta$, where $T_{\rm EW}\approx 140\,{\rm GeV}$
is the temperature
of sphaleron freeze-out that therefore occurs when $z=1$.
In this parametrisation, the kinetic equations read
\begin{subequations}
\label{kineq:z}
\begin{align}
\label{kineq:N}
\frac{d}{dz}n_{Ni}
&=-\sum_a Y_{ia} Y_{ai}^\dagger\frac{\Gamma_{\rm av}}{T_{\rm EW}}(n_{Ni}-n_N^{\rm eq})\,,
\\
\frac{d}{dz}\frac{q_{\ell aa}}{s}&= g_w S_{aa}/(s T_{\rm EW})\,.
\end{align}
\end{subequations}
(Recall the conformal rescaling of the masses in $S_{aa}$.)
The entropy density is given by
\begin{align}
s=\frac{2\pi^2}{45}g_\star T^3\,.
\end{align}

Within the kinetic equations~(\ref{kineq:eta},\ref{kineq:z}), we have not included
washout terms, because it turns out that these are negligible at the time where
the lepton asymmetry is produced. Washout is however of crucial importance close
to the Electroweak scale, where the different rates for the particular active
lepton flavours can effectively turn the flavoured asymmetries into a
baryon asymmetry, as we discuss in Section~\ref{sec:par:examples}.

For the source terms~(\ref{Source:approximate}), we have assumed that
the density of the sterile neutrinos is very small compared to
their equilibrium density at the time when the main contributions
to the asymmetry are generated. Using Eq.~(\ref{kineq:N}),
it is useful to estimate the size of the Yukawa couplings for
which equilibration occurs at $T_{\rm EW}$ as
\begin{align}
\label{bound:Y}
\sum_a Y_{ia} Y_{ai}^\dagger\frac{\Gamma_{\rm av}}{H_{\rm EW}}
\sim 1
\;\Rightarrow\;
\sum_a Y_{ia} Y_{ai}^\dagger
\sim
\frac{H_{\rm EW}}{\Gamma_{\rm av}}\approx 17\times \frac{T_{\rm EW}}{\gamma_{\rm av}m_{\rm Pl}}\approx2.0\times 10^{-16}/\gamma_{\rm av}\,,
\end{align}
where we have set $T_{\rm EW}\approx140\,{\rm GeV}$, and
$H_{\rm EW}$ is the Hubble rate at the temperature $T_{\rm EW}$.
With the same parametrisations and approximations, the
source terms are estimated as
\begin{subequations}
\begin{align}
\label{S:LNV}
g_w S_{aa}^{\rm LNV} /(s T_{\rm EW})&\approx
-\sum\limits_{\overset{c}{j\not=i}}
{\rm i}\frac{Y^\dagger_{ai}Y^*_{ic}Y^t_{cj}Y_{ja}
-Y^t_{ai}Y_{ic}Y^\dagger_{cj}Y^*_{ja}}{M_{ii}^2-M_{jj}^2}
M_{ii}M_{jj}\frac{m_{\rm Pl}}{T_{\rm EW}} \gamma_{\rm av}^2\times 5.3\times 10^{-5}\,,
\\
\label{S:PF}
g_w S_{aa}^{\rm PF}/(s T_{\rm EW})
&\approx
-\sum\limits_{\overset{c}{j\not=i}}
{\rm i}\frac{Y^\dagger_{ai}Y_{ic}Y^\dagger_{cj}Y_{ja}
-Y^t_{ai}Y^*_{ic}Y^t_{cj}Y^*_{ja}}{M_{ii}^2-M_{jj}^2}
\frac{m_{\rm Pl}T_{\rm EW}}{z^2} \gamma_{\rm av}^2 \times 7.3\times 10^{-4}\,.
\end{align}
\end{subequations}
Note that in these expressions (in particular for
the purely flavoured term), the masses are understood
as physical masses, {\it i.e.} without conformal rescaling.
The singularity in the purely flavoured contribution for
$z\to 0$ is regulated by the requirement that the
mass splitting
$a(\eta)^2(M_{\rm ii}^2-M_{\rm jj}^2)\gsim
\underset{i}{\rm max}\sum_aY_{ia}Y^\dagger_{ai}|\mathbf k|\Gamma_{\rm av}$,
what is equivalent to our assumption that many sterile
flavour oscillations occur within one decay time. For the
typical momentum scale with $|\mathbf k|\sim T$, this translates
to
\begin{align}
\label{zbound:damp}
z^2\gsim\frac{T_{\rm EW}^2}{|M_{ii}^2-M_{jj}^2|}
\underset{i}{\rm max}\sum_aY_{ia}Y^\dagger_{ai} \gamma_{\rm av}\,.
\end{align}
We denote the temperature, at which this bound is saturated by $T_{\rm res}$.
Below $T_{\rm res}$,
the mass difference dominates over the damping rates $\Gamma_i$, such that
expression~(\ref{Source:flavoured:2})
for the resonantly enhanced source is applicable.
In addition, in order our approximation to be valid,
we must require that the sterile neutrinos have performed at least one
oscillation before the production of the asymmetry begins
($a^2|M_{ii}^2-M_{jj}^2|\eta/(2 T)\gsim 2\pi$), what leads to
\begin{align}
\label{zbound:osc}
z^3\gsim8\pi\sqrt{\frac{\pi^3 g_\star}{45}}\frac{T_{\rm EW}^3}{m_{\rm Pl}|M_{ii}^2-M_{jj}^2|}\;
\Rightarrow\;z\gsim6.0\times\left(\frac{T_{\rm EW}^3}{m_{\rm Pl}|M_{ii}^2-M_{jj}^2|}\right)^{\frac 13}\,.
\end{align}
We denote the temperature at which this latter bound is saturated by
$T_{\rm osc}$, which has been
identified earlier in Ref.~\cite{Akhmedov:1998qx} as the temperature,
at which the most relevant contributions to the flavoured
asymmetries emerge. In Ref.~\cite{Garbrecht:2011aw}, it has been demonstrated
that provided the bounds~(\ref{zbound:damp}) and~(\ref{zbound:osc}) are met,
using the source term~(\ref{Source:flavoured:2}) is approximately equivalent
to solving for the time-evolution of the distribution functions
$\delta f_{0hij}$ of the sterile neutrinos.
The kinetic equation for the the time-evolution
of $\delta f_{0hij}$ is obviously equivalent to the canonical time-evolution
of a density matrix, that is commonly employed to describe neutrino
oscillations~\cite{Dolgov:1980cq,Barbieri:1990vx,Enqvist:1990ad,Enqvist:1991qj,Sigl:1992fn}
and that is used in Refs.~\cite{Asaka:2005pn,Shaposhnikov:2008pf,Canetti:2010aw}
in order to compute the lepton asymmetry. Therefore, we are considering here
the same underlying mechanism as in Refs.~\cite{Asaka:2005pn,Shaposhnikov:2008pf,Canetti:2010aw}. However, our result for the
flavoured asymmetries, Eq.~(\ref{flavoured:asymmetries}) below, takes a simple
analytic form that
is suitable for fast
parametric surveys and that thus allows to identify viable scenarios
for Leptogenesis with non-degenerate ${\rm GeV}$-scale sterile neutrinos. Moreover,
the source term in the form~(\ref{Source:flavoured:2}) clarifies the relation of
the present scenario to Leptogenesis from heavier sterile neutrinos,
that are more commonly discussed in the literature~\cite{Fukugita:1986hr,DiBari:2012fz,Davidson:2002qv,Buchmuller:2002rq,Endoh:2003mz,Pilaftsis:2005rv,Abada:2006fw,Nardi:2006fx,DiBari:2005st,Engelhard:2006yg,Antusch:2010ms,Beneke:2010dz,Covi:1996wh,Flanz:1996fb,Pilaftsis:1997dr,Pilaftsis:1997jf,Pilaftsis:2003gt,Beneke:2010wd,Pilaftsis:2004xx,Asaka:2008bj,Blanchet:2009kk,Racker:2012vw}.

In order not to substantially overestimate the asymmetry,
one should take those of the bounds~(\ref{zbound:damp}) and~(\ref{zbound:osc})
that requires the larger value for $z$.
Integrating Eq.~(\ref{S:PF}) from
the bound~(\ref{zbound:damp}) to $z=1$, we obtain
\begin{align}
\frac{q_{\ell a}}{s}
\equiv
\frac{q_{\ell aa}^{\rm PF}}{s}
\approx
-\frac{1}{\sqrt{\underset{i}{\rm max}\sum_aY_{ia}Y^\dagger_{ai}}}
\sum\limits_{\overset{c}{j\not=i}}
{\rm i}
\frac
{
Y^\dagger_{ai}Y_{ic}Y^\dagger_{cj}Y_{ja}
-Y^t_{ai}Y^*_{ic}Y^t_{cj}Y^*_{ja}
}
{
{\rm sign}(M_{ii}^2-M_{jj}^2)
}
\frac{m_{\rm Pl}\times 5.3\times10^{-5}\gamma_{\rm av}^{\frac32}}{\sqrt{|M_{ii}^2-M_{jj}^2|}}\,,
\end{align}
whereas integrating from the bound~(\ref{zbound:osc})
to $z=1$ leads to
\begin{align}
\label{flavoured:asymmetries}
\frac{q_{\ell a}}{s}\equiv\frac{q_{\ell aa}^{\rm PF}}{s}
\approx
-\sum\limits_{\overset{c}{j\not=i}}
{\rm i}
\frac
{
Y^\dagger_{ai}Y_{ic}Y^\dagger_{cj}Y_{ja}
-Y^t_{ai}Y^*_{ic}Y^t_{cj}Y^*_{ja}
}
{
{\rm sign}(M_{ii}^2-M_{jj}^2)
}
\left(\frac{m_{\rm Pl}^2}{|M_{ii}^2-M_{jj}^2|}\right)^{\frac 23}\times 1.2\times 10^{-4}\gamma_{\rm av}^2\,.
\end{align}
Since we consider in the following parametric
examples where in Eqs.~(\ref{zbound:damp}) and (\ref{zbound:osc})
$z\ll 1$, it is justified that we have approximated the
integrals over $z$ by the contribution from the lower boundary.
Furthermore,
for our parametric examples in Section~\ref{sec:par:examples}, it turns out that
$T_{\rm osc}<T_{\rm res}$, and consequently, we use Eq.~(\ref{flavoured:asymmetries})
for the flavoured asymmetries. Note that this expression differs in various
aspects from the estimate of the asymmetry given in Ref.~\cite{Akhmedov:1998qx}. On the other hand, the parametric dependences
agree with what is found in Ref.~\cite{Asaka:2005pn}, while the
numerical prefactor here is enhanced. This may perhaps
partly be explained because the source for the asymmetry
is proportional to $\mathbf k^2$, such that it is largely
enhanced for larger momenta, while the estimates in Ref.~\cite{Asaka:2005pn} rely on the assumption that the relevant contributions
from phase space occur for $|\mathbf k|\sim T$.

The present calculation relies on a systematic derivation of the asymmetry
based on the CTP formalism.
Moreover, we find that the dependence of the asymmetry
on the Yukawa couplings agrees with what is reported for conventional
scenarios of flavoured Leptogenesis~\cite{Covi:1996wh,Endoh:2003mz,Nardi:2006fx};
and the
underlying source term~(\ref{Source:flavoured:2}) emerges as a generalisation
of the earlier
results~\cite{Covi:1996wh,Endoh:2003mz,Nardi:2006fx} to finite-temperature backgrounds.
As a consequence, we find that the asymmetry is suppressed by the factor
$\gamma_{\rm av}^2$, whereas the suppression in Ref.~\cite{Akhmedov:1998qx}
is only by a factor of $\gamma_{\rm av}$ (to be identified with $\sin\phi$
in that paper). Moreover, there is a different dependence on the difference
between the squared masses and on $m_{\rm Pl}$.

The lepton number-violating contribution to the asymmetry can directly
be inferred from integrating the source~(\ref{S:LNV}) over
$z$ from $0$ to $1$. In particular, we may approximate the lower
boundary by $0$, as the integrand is independent of $z$
and therefore finite for $z\to 0$. One should remark however
that this procedure has to be modified when the RH
neutrino distributions cannot be approximated as vanishing
close to $z=1$, as it is done in Eq.~(\ref{Source:approximate}).
Using the estimate~(\ref{bound:Y}) and Eq.~(\ref{S:LNV}),
in order to achieve a lepton asymmetry of order
$2.5\times 10^{-10}$ from the lepton-number violating contribution, we
require that $M_{ii}^2/|M_{ii}^2-M_{jj}|^2\gsim 3.5 \times 10^8$
(where $M_{ii}^2\approx M_{jj}^2$). This is to be compared with the
non-degenerate mass parameters
that can explain Baryogenesis via flavoured Leptogenesis, that are given for specific
points in parameter space in the following Section.
Parametrically, the relative enhancement of the flavoured asymmetries
in the ARS mechanism over the lepton-number violating asymmetries,
that are of importance in many other scenarios for Leptogenesis
and that underlie the mass bounds in Refs.~\cite{Davidson:2002qv,Buchmuller:2002rq}, can be seen when comparing
Eqs.~(\ref{S:LNV}) and~(\ref{flavoured:asymmetries}).

\section{Flavoured Leptogenesis without Mass Degeneracy}
\label{sec:par:examples}

It is already clear at this point that the purely flavoured asymmetries
are generically much enhanced when compared to the lepton-number
violating asymmetry. In the present Section, we investigate
typical slices of parameter space on which
a lepton flavour asymmetry of order of the observed baryon asymmetry
can emerge without or with only a moderate degeneracy of
the sterile neutrino masses.
The main requirements are that
the parameters of the model must be consistent with the
oscillations of active neutrinos
(see {\it e.g.} Ref.~\cite{Fogli:2012ua} for the most recently determined parameters)
and the lepton flavour asymmetries
must be partially converted into a baryon asymmetry due to sphaleron
processes. The latter requirement can be met due to the
different couplings of the active lepton flavours to the sterile
neutrinos, that lead to different washout rates
before sphaleron freeze-out. In particular, we find that it is
possible that the washout for the flavours $\mu$ and $\tau$ is
maximal, whereas the flavour $e$ is only partially washed out.
Therefore, the asymmetry~(\ref{flavoured:asymmetries}) in the flavour $e$
can be almost maximally converted into a baryon asymmetry, with a suppression factor of order unity due to partial
washout and cancellations, that we discuss toward the end of this
Section.

In order to obtain parameters that are consistent with the observed neutrino
oscillations,
we use the Casas-Ibarra parametrisation for the Yukawa
couplings~\cite{Casas:2001sr}, 
\begin{equation}
\label{CasasIbarraDef}
%F=
Y^\dagger=U_\nu \sqrt{m_\nu^{\rm diag}} \mathcal{R} \sqrt{M_M} \frac{\sqrt 2}{v},
\end{equation}
where $m_{\nu}^{\rm diag}={\rm diag}(m_1,m_2,m_3)$ is the diagonal mass matrix
of the active leptons,
$M={\rm diag}(M_{11},M_{22},M_{33})$
and $v=246\,{\rm GeV}$.
The PMNS matrix $U_{\nu}$ can be parametrised as
\begin{equation}
U_\nu=V^{(23)}U_{\delta}V^{(13)}U_{-\delta}V^{(12)}{\rm diag}(e^{i\alpha_1 /2},e^{i\alpha_2 /2},1)
\end{equation}
with $U_{\pm\delta}={\rm diag}(e^{\mp i\delta/2},1,e^{\pm i\delta/2})$ and 
the non-zero entries of the matrices $V$ are given by
\begin{eqnarray}
%	U_{\nu}&=&
V^{(ij)}_{ii}=V^{(ij)}_{jj}=c_{ij}\,,\quad
V^{(ij)}_{ij}=s_{ij}\,,\quad
V^{(ij)}_{ji}=-s_{ij}\,,\quad
V^{(ij)}_ {kk}\underset{k\not=i,j}{=}1\,,
V^{(ij)}_ {kl}\underset{k\not=i,j}{=}1\,,
\end{eqnarray}
where $c_{ij}$ and $s_{ij}$ stand for $\cos(\uptheta_{ij})$ and $\sin(\uptheta_{ij})$, respectively, $\uptheta_{ij}$ are the mixing angles
of the active leptons, and $\alpha_1$, $\alpha_2$ and $\delta$ are the $CP$-violating phases. The matrix ${\cal R}$ must satisfy
${\cal R}^t{\cal R}=1$.
%For normal hierarchy the Yukawa matrix $Y$ only depends effectively
%on the phases $\alpha_2$ and $\delta$, for the inverted hierarchy,
%it depends on $\delta$ and the difference $\alpha_1-\alpha_2$.
%This is because $N_1$ has no measurable effect on neutrino masses due to $M_1\ll M_{2,3}$.
It can be parametrised by the complex
angles $\omega_{ij}$ and the matrices ${\cal R}^{(ij)}$
with the non-zero entries
\begin{align}
{\cal R}^{(ij)}_{ii}&={\cal R}^{(ij)}_{jj}=\cos\omega_{ij}\,,\quad
{\cal R}^{(ij)}_{ij}=\sin\omega_{ij}\,,\quad
{\cal R}^{(ij)}_{ji}=-\sin\omega_{ij}\,,\quad
{\cal R}^{(ij)}_{ii}\underset{k\not=i,l}{=}1\,,
\\\notag
{\cal R}&={\cal R}^{(23)}{\cal R}^{(13)}{\cal R}^{(12)}\,.
\end{align}

The washout rates for the particular lepton flavours are given by
\begin{align}
\label{washout}
\Gamma_a=\frac{1}{g_w}\sum\limits_i Y_{ai}^\dagger Y_{ia} \gamma_{\rm av} T\,,
\end{align}
where $a=e,\mu,\tau$. As stated above, Ref.~\cite{Besak:2012qm} suggests that
$\gamma_{\rm av}=5\times10^{-3}$ at Electroweak temperatures.

\begin{table}[th!]
\begin{center}
\begin{tabular}{cc}
\begin{tabular}{|r|r|r|}
\hline
& Scenario I & Scenario II\\
\hline
$m_1$ & $0$ & $2.5\,{\rm meV}$\\
$m_2$ & $8.7\,{\rm meV}$ & $9.1\,{\rm meV}$\\
$m_3$ & $49\,{\rm meV}$ & $49\,{\rm meV}$\\
$s_{12}$ & $0.55$ & $0.55$\\
$s_{23}$ & $0.63$ & $0.63$\\
$s_{13}$ & $0.16$ & $0.16$\\
$\delta$ & $-\pi/4$ & $\pi$\\
$\alpha_1$ & $0$ & $-\pi$\\
$\alpha_2$ & $-\pi/2$ & $\pi$\\
$\omega_{12}$ & $1.0+2.6{\rm i}$ & $-1.0+1.5{\rm i}$\\
$\omega_{13}$ & $0.9+2.7{\rm i}$ & $0.5+2.6{\rm i}$\\
$M_1$ & $3.6\,{\rm GeV}$ & $1.0\,{\rm GeV}$\\
$M_2$ & $4.0\,{\rm GeV}$ & $2.0\,{\rm GeV}$\\
$M_3$ & $4.4\,{\rm GeV}$ & $3.0\,{\rm GeV}$\\
\hline
\end{tabular}
&
\begin{tabular}{|r|r|r|}
\hline
& Scenario I & Scenario II\\
\hline
$\omega_{23}$ & $0.0-1.2{\rm i}$ & $\pi-2.4{\rm i}$\\
$q_{e}/(s\times2.5\times 10^{-10})$ &  $-6.7$ & $-8.3$\\
$q_{\mu}/(s\times2.5\times 10^{-10})$ & $-31.8$ & $32.0$\\
$q_{\tau}/(s\times2.5\times 10^{-10})$ & $-25.1$ & $-23.7$\\
$\Gamma_{e}/H_{\rm EW}$ & $0.64$ & $0.59$\\
$\Gamma_{\mu}/H_{\rm EW}$ & $290$ & $420$\\
$\Gamma_{\tau}/H_{\rm EW}$ & $930$ & $150$\\
$T_{\rm osc}$ & $5\times 10^5\,{\rm GeV}$& $5\times 10^5\,{\rm GeV}$\\
$T_{\rm res}$ & $3.2\times 10^8\,{\rm GeV}$& $ 4.4 \times 10^8\,{\rm GeV}$\\
%$\Gamma^{\rm hfl}_{e}/H_{\rm EW}$ & $0.33$ & $0.13$\\
%$\Gamma^{\rm hfl}_{\mu}/H_{\rm EW}$ & $42$ & $43.7$\\
%$\Gamma^{\rm hfl}_{\tau}/H_{\rm EW}$ & $88$ & $106.4$\\
\hline
\end{tabular}
\\
\\
(A) & (B)
\end{tabular}
\end{center}
\vskip-.4cm
\caption{\label{table:scenarios}
In Table~(A), two sets of parameters that lead to viable Baryogenesis via
Leptogenesis in conjunction with the values for
$\omega_{23}$ in Table~(B). In addition,
in Table~(B), derived parameters relevant for Leptogenesis in these
Scenarios.}
\end{table}

For definiteness, we consider the case of a normal hierarchy
and two Scenarios given by the parameters in Table~\ref{table:scenarios}(A).
For these choices of parameters, $|M_{ii}^2-M_{jj}^2|\gsim 3\,{\rm GeV}^2$,
such that the relation~(\ref{zbound:osc}) implies that
$z\gsim 9 \times 10^{-5}$. Using $T=T_{\rm EW}/z$, this means that the lepton
asymmetry emerges at temperatures of about $T=T_{\rm osc}=5 \times10^5\,{\rm GeV}$.
For consistency, it should be checked that $T_{\rm osc}<T_{\rm res}$, which
is indeed the case for the points in parameter space given in Table~\ref{table:scenarios}.
The tabulated rates of Ref.~\cite{Besak:2012qm} therefore imply that
we take $\gamma_{\rm av}=3\times 10^{-3}$ for obtaining the values of the
flavour asymmetries~(\ref{flavoured:asymmetries}).
With all other parameters fixed, we investigate the slice of parameter space
from varying $\omega_{12}$. 

\begin{figure}[t!]
\begin{center}
\begin{tabular}{cc}
\hspace{-.4cm}
\epsfig{file=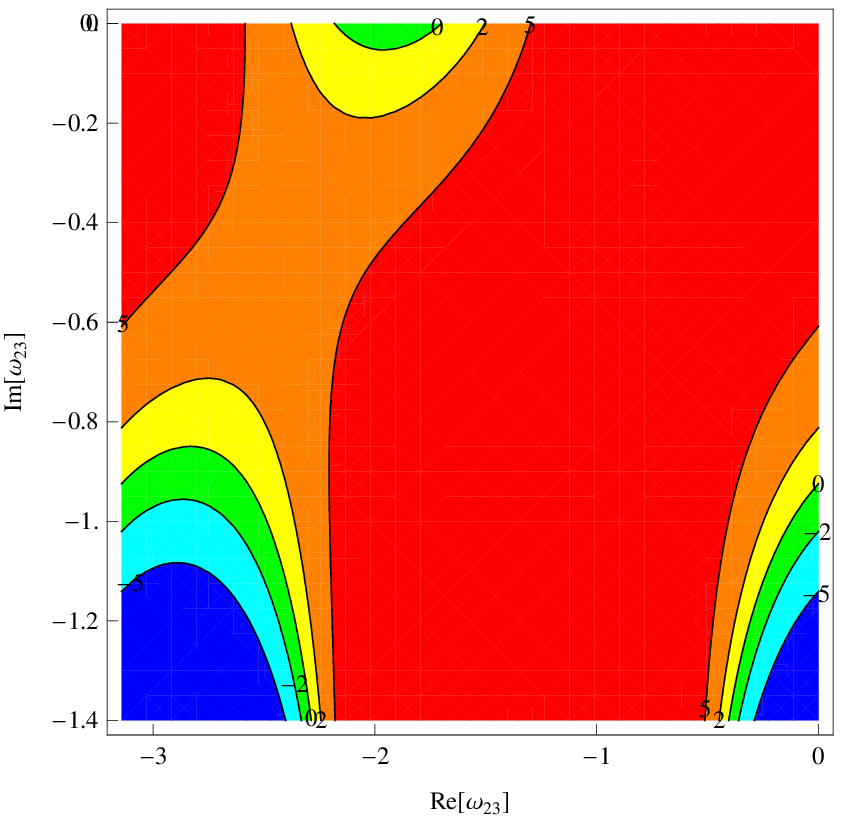,width=7.5cm}
&
\epsfig{file=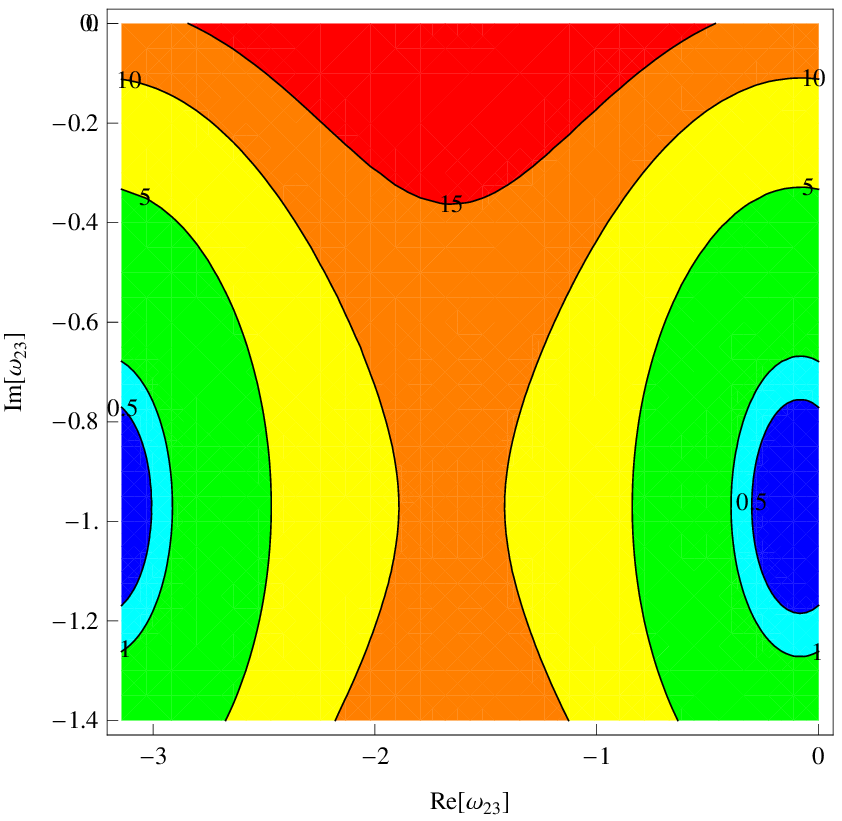,width=7.5cm}
\\
(A)&(B)
\end{tabular}
\end{center}
\vskip-.4cm
\caption{
\label{fig:scI}
In~(A) the asymmetry $q_{e}/(s\times2.5\times 10^{-10})$
and in~(B) the washout $\Gamma_e/H_{\rm EW}$
of the electron-flavour in Scenario~I.
The result has the period $\pi$ in ${\rm Re}[\omega_{23}]$.
Successful Baryogenesis via Leptogenesis occurs when large negative
asymmetries coincide with small washout rates, as in the bottom corners
of the diagrams.
}
\end{figure}

\begin{figure}[t!]
\begin{center}
\begin{tabular}{cc}
\hspace{-.4cm}
\epsfig{file=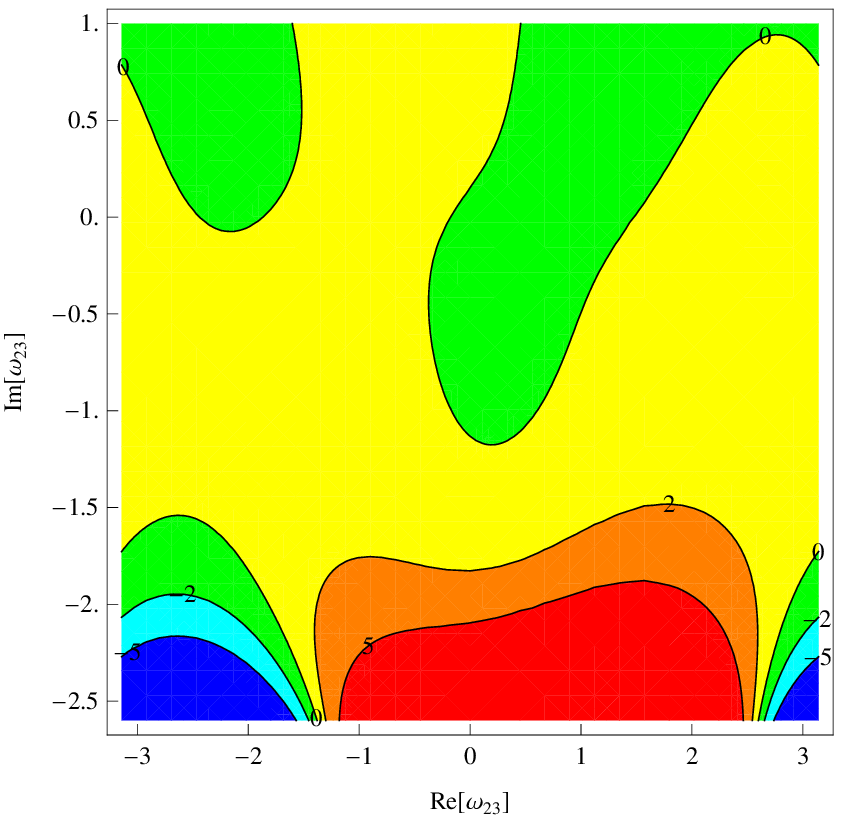,width=7.5cm}
%\hspace{.5cm}
&
%\hspace{.5cm}
\epsfig{file=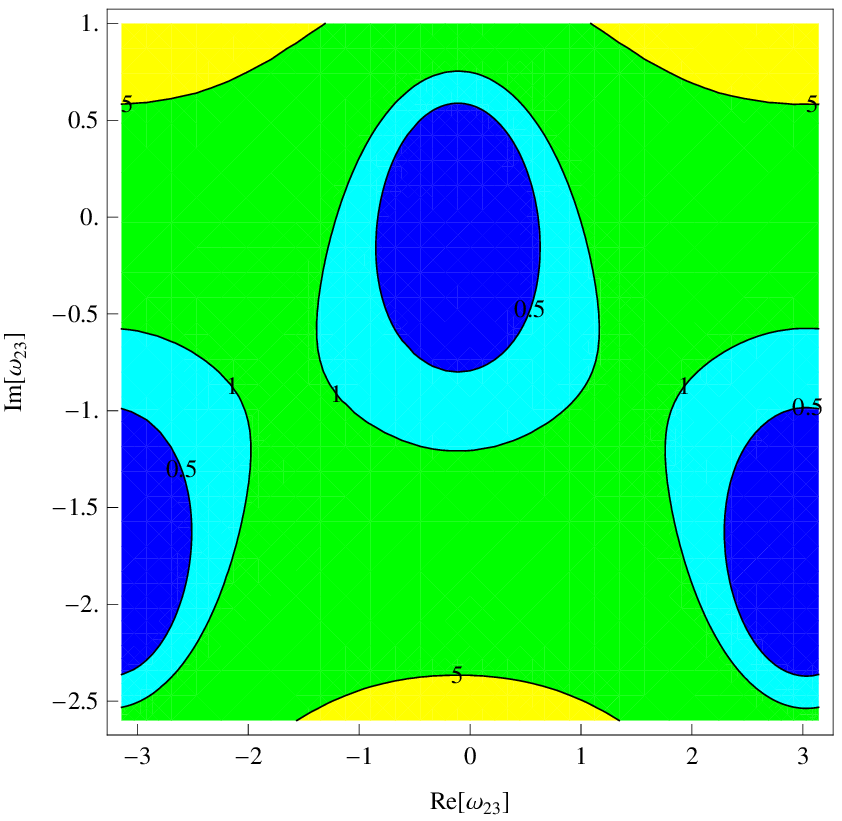,width=7.5cm}
\\
(A)&(B)
\end{tabular}
\end{center}
\vskip-.4cm
\caption{
\label{fig:scII}
In~(A) the asymmetry $q_{e}/(s\times2.5\times 10^{-10})$
and in~(B) the washout $\Gamma_e/H_{\rm EW}$
of the electron-flavour in Scenario~II.
The result has the period $2\pi$ in ${\rm Re}[\omega_{23}]$.
Successful Baryogenesis via Leptogenesis occurs when large negative
asymmetries coincide with small washout rates, as in the bottom corners
of the diagrams.
}
\end{figure}

In both scenarios [for $\omega_{23}$ as given in Table~\ref{table:scenarios}(B)],
the sterile neutrinos are in thermal equilibrium and the washout of the $\mu$
and $\tau$ flavours is generically large compared to
the Hubble rate at the Electroweak scale, $H_{\rm EW}$. 
In contrast, the $e$ flavour has not reached chemical equilibrium because its mixing with sterile neutrinos is suppressed with respect to $\mu$ and $\tau$. This possibility has previously been studied in different context~\cite{Asaka:2011pb}.
Due to the large mixing with the $\mu$ and $\tau$ flavours, the sterile neutrinos decay sufficiently long before Big Bang Nucleosynthesis not to affect the predicted light element abundances, {\it cf.} Refs.~\cite{Ruchayskiy:2012si,Canetti:2012vf}.
At the same time, the scenarios [with $\omega_{23}$ as in Table~\ref{table:scenarios}(B)] are consistent with the negative results of direct searches in the past, which have been interpreted in the context of seesaw models in~\cite{Asaka:2011pb,Ruchayskiy:2011aa,Canetti:2012vf}.
In Figures~\ref{fig:scI}(A) and~\ref{fig:scII}(A), we show the asymmetry in the flavour $e$
normalised to the observed asymmetry (already accounting for the sphaleron
conversion factor). Figures~\ref{fig:scI}(B)
and~\ref{fig:scII}(B) display the washout rates of the flavour $e$
normalised to the Hubble expansion rate at sphaleron freeze-out.
From the asymmetry and the washout rates, we can identify values
for $\omega_{12}$, where Baryogenesis via Leptogenesis is viable. These
values, for which a large asymmetry $q_e$ is produced and at the same time
washout for the $e$-flavour
is small, are given in Table~\ref{table:scenarios}(B).

In particular, from the values of $\Gamma_\mu$ and $\Gamma_\tau$, we deduce that
these are equilibrating with the sterile neutrinos before sphaleron
freeze-out. It can also be verified that
both, $\mu$ and $\tau$, equilibrate with all three individual $N_i$ before
sphaleron freeze-out.
Accounting for the two ${\rm SU}(2)$ degrees
of freedom and the fact that there are three sterile neutrinos,
only 3/7 of the asymmetry in $\mu$ and $\tau$ is transferred to the sterile sector.
Since the sum of the flavoured asymmetries is vanishing, this implies that
4/7 of the asymmetry $q_e/s$ are cancelled by $q_\mu/s$ and $q_\tau/s$.
More precisely, the large interaction rates of $\mu$ and $\tau$
imply the equilibrium conditions
$\mu_{\mu}=\mu_{\tau}=\mu_{N 1,2,3}$ for the chemical potentials of
the particular particles. When we denote by $q_{\mu,\tau}^{\rm W}$ the
charge density in the $\mu$ and $\tau$-leptons after equilibration
with $N_{1,2,3}$ (washout), and accordingly, we define
$q_{N 1,2,3}^{\rm W}$,
this gives the relation
$g_w q^{\rm W}_{\mu}=g_w q^{\rm W}_{\tau}=q^{\rm W}_{N 1,2,3}$. Lepton number
conservation
implies
$q^{\rm W}_{\mu}+q^{\rm W}_{\tau}+q^{\rm W}_{N 1}+q^{\rm W}_{N 2}+q^{\rm W}_{N 3}=q_\mu+q_\tau=-q_e$. With $g_w=2$, the latter two equations imply that
$q^{\rm W}_{\mu}+q^{\rm W}_{\tau}=-(4/7)q_e$ is the part of the asymmetry,
that is cancelled before sphaleron freeze-out.
Furthermore, from the values of $\Gamma_e$, we estimate that about half of
the asymmetry in this flavour is transferred to the sterile sector, where
it annihilates with the asymmetries in $\mu$ and $\tau$. Consequently,
the baryon asymmetry can be estimated as $-3/14\times0.35\times q_e/s$
($0.35$ being the sphaleron-conversion factor),
and the parametric configurations
in Table~\ref{table:scenarios} may therefore account for the
observed baryon asymmetry and for the observed neutrino oscillations.

We emphasise that the Scenarios~I and~II correspond to particular points in
parameter space, and that the present analysis does not quantify how generically
successful Baryogenesis via flavoured Leptogenesis for ${\rm GeV}$-scale
sterile neutrinos can occur. Nonetheless, the present examples demonstrate that the
order $10^8$ mass degeneracy, that is necessary for generating
a sufficient amount of lepton-number violating asymmetries, can be largely relaxed
for the flavoured scenarios.

\section{Discussion and Conclusions}
\label{sec:disc}

We have demonstrated that for ${\rm GeV}$-scale sterile neutrinos
without a mass degeneracy,
flavoured Leptogenesis can be a viable mechanism to generate the
baryon asymmetry of the Universe. 
The simple analytic expression~(\ref{flavoured:asymmetries})
has been used to identify
viable points in parameter space with no mass-degeneracy,
which is in contrast
to scenarios that have been discussed earlier and that rely on sterile
neutrinos close to the Electroweak scale or
below~\cite{Asaka:2005pn,Pilaftsis:2005rv,Shaposhnikov:2008pf,Canetti:2010aw}.
The key ingredient to the variant of the ARS mechanism that
is studied here is the fact that
the flavoured asymmetry is for our parameters most effectively generated at
$T_{\rm osc}\approx 5\times 10^5\,{\rm GeV}\gg M_{ii}$, where thermal effects can considerably enhance it. For higher temperatures,
the mass difference is small compared to the Hubble rate and there is no time for the neutrinos to oscillate (or, in other
words, for off-diagonal correlations to build up),
what leads to a suppression. Compared to lower temperatures, the production
at $T_{\rm osc}\approx 5\times 10^5{\rm GeV}$ is larger due to
the resonant enhancement factor~(\ref{enhancement:flavoured})
and the neutrino production rate~(\ref{gamma:N}). On the other hand,
an effective suppression results from the smaller time scale
associated with $T_{\rm osc}$, $H^{-1}\sim m_{\rm Pl}^2/T_{\rm osc}$.
The temperature $T_{\rm osc}$ has been identified earlier as
a favourable instant for the creation of lepton asymmetries in Ref~\cite{Akhmedov:1998qx}.
In Eq.~(\ref{Source:flavoured:2}),
we have generalised the $CP$-violating source term of Refs.~\cite{Beneke:2010wd,Garbrecht:2011aw}, such that
it includes the possibility of non-degenerate
active lepton-flavours. Comparing the lepton-number violating with
the purely flavoured contribution, we observe that the latter does not exhibit
the suppression factor $M_{ii} M_{jj}$ that is associated with the
lepton-number violating helicity flip. The Eqs.~(\ref{Sigma:T0})
and~(\ref{Sigma:Tlarge}) are a consistency check and further illustrate
the importance of finite-temperature effects.
In particular, they elucidate how the ARS scenario circumvents the
mass bounds~\cite{Davidson:2002qv,Buchmuller:2002rq} that apply to the
total lepton-number violation or when finite temperature effects
are negligible.
A similar
type of enhancement of the flavoured asymmetries, that
relies on the ratio between the square of the temperature
and the difference of the squares of the masses of Higgs bosons has been
proposed in Ref~\cite{Garbrecht:2012qv}.

We have used here nonequilibrium Quantum Field Theory
methods, in particular the CTP formalism,
in order to calculate the asymmetries.
In Refs.~\cite{Akhmedov:1998qx,Asaka:2005pn}, the canonical
time-evolution of a density matrix is employed for this purpose.
Our result for the flavoured asymmetry differs from
Ref.~\cite{Akhmedov:1998qx}, but it parametrically agrees
with Ref.~\cite{Asaka:2005pn}. Moreover, our flavoured asymmetries
appear as a consistent generalisation of
the results from Refs.~\cite{Covi:1996wh,Endoh:2003mz,Nardi:2006fx},
where the decay asymmetries are calculated from $S$-matrix elements.
The present work therefore
exhibits the close relation of the ARS mechanism to
the more conventional scenarios of flavoured thermal Leptogenesis.

Phenomenologically, we have demonstrated the viability of the mechanism
with non-degenerate masses by
calculating the lepton asymmetry for
the parameters given in Table~\ref{table:scenarios}.
That we do
not need to require a mass degeneracy here is due to the fact that we
find parameters, for which it is possible that the asymmetry in the flavour
$e$ is only weakly washed out, while $\mu$ and $\tau$
already equilibrate
with the sterile neutrinos above $T_{\rm EW}$.
More systematic studies of the parameter space should however be of interest.
The simple analytic formula~(\ref{flavoured:asymmetries}) for the
asymmetry may be of particular use for such analyses.
For the scenarios in Table~\ref{table:scenarios},
we find that the mass
degeneracy of the sterile neutrinos
can be completely alleviated or at least be largely relaxed
compared to what is stated {\it e.g.} in
Refs.~\cite{Asaka:2005pn,Pilaftsis:2005rv,Shaposhnikov:2008pf,Canetti:2010aw}.
However, it should be noticed that
the parameters $\omega_{ij}$ have no direct physical interpretation
and that the moderately large values we consider correspond to
cancellations of different contributions to the masses of active leptons.
In other words, the Yukawa couplings are somewhat larger than
for most of the points in parameter space for the seesaw
mechanism.
In the future, it will therefore be interesting to pursue possibilities that
either do not require such cancellations or explain these:
\begin{itemize}
\item
The estimate of $\Gamma^{\rm av}$, that is taken from Ref.~\cite{Besak:2012qm},
relies on the simple extrapolation of the Standard Model with additional
sterile neutrinos to high temperatures, such that the main production
channels for the sterile neutrinos are via scatterings of ${\rm SU}(2)_{\rm L}$
and ${\rm U}(1)_Y$ gauge bosons. In many extensions of the Standard Model,
interactions with an extended Higgs sector, with the Dark Matter sector or
larger couplings are encountered at higher energies.
This would lead to an enhancement
of $\Gamma^{\rm av}$ and would be of importance for the asymmetry,
that, through the source term~(\ref{Source:flavoured:2}), depends
quadratically on $\Gamma^{\rm av}$.
\item
Larger Yukawa couplings for fixed masses of the active neutrinos can
be achieved in models with several Higgs doublets. In order to
avoid flavour problems, an additional leptoholic Higgs,
that exclusively couples to leptons, may be suitable.
\item
Models with an approximate lepton-number conservation can typically
accommodate for largely enhanced Yukawa couplings~\cite{Branco:1988ex,GonzalezGarcia:1988rw,Pilaftsis:2004xx,Pilaftsis:2005rv,Shaposhnikov:2006nn,Asaka:2008bj,Blanchet:2009kk,Ibarra:2011xn,Racker:2012vw}.
It would be interesting to study the implications of the
thermal effects in the source
term~(\ref{Source:flavoured:2}) for these scenarios.
\item Sterile neutrinos with masses of a few GeV can in principle be found in laboratory experiments \cite{Gorbunov:2007ak}.
It has been shown in \cite{Canetti:2010aw,Canetti:2012vf} that this possibility certainly exists when the BAU is generated from mass-degenerate sterile neutrinos.
For the exemplary parts of the parameter space we study, the active-sterile mixing is too small to search for them in existing or near-future experimental facilities. Larger mixings are essentially forbidden because at least one active flavours must be out of chemical equilibrium at $\sim T_{EW}$ to avoid a complete washout of the flavour asymmetries. 
This is despite the fact that we consider a situation in which the mixings of two active flavours are large compared to the third one. It remains to be clarified in systematic scans of the higher-dimensional parameter-space for models with several right handed neutrinos whether there are regions in which this suppression is sufficiently strong such
that sterile neutrinos without mass degeneracy that can explain the BAU may at the same time be directly experimentally accessible. 
\end{itemize}
These are intriguing prospects, whether there is a Higgs sector
as in Standard Model or in an extended form.
Even with non-degenerate masses at the ${\rm GeV}$-scale, sterile neutrinos
thus provide a plausible possible
explanation for the baryon asymmetry of the Universe.

\subsection*{Acknowledgements}
We would like to thank Alejandro Ibarra for useful remarks on large
imaginary parts of $\omega_{ij}$.
This work is supported by the Gottfried Wilhelm Leibniz programme
of the Deutsche Forschungsgemeinschaft.

\end{document}